\documentclass[A4, 12pt, review]{elsarticle}

\usepackage{multirow}
\usepackage{subfigure}
\usepackage{amsmath}
\usepackage{xcolor,colortbl}
\definecolor{pearl}{rgb}{0.94, 0.92, 0.84}
\definecolor{lightmauve}{rgb}{0.86, 0.82, 1.0}
\usepackage{makecell}
\usepackage{arydshln}
\usepackage{color,soul}
\usepackage{gensymb}
\usepackage{graphicx}
\usepackage{times}
\definecolor{black}{rgb}{0.0, 0.0, 0.0}
{}
{}
{}
\usepackage{geometry}
\usepackage{lineno,hyperref}

\journal{arXiv}

\begin{document}

\begin{frontmatter}

\title{\textbf{Investigation of Genomic Effect of Zirconium Oxide Nanoparticles in \textit{Escherichia coli} Bacteria}}

\author[]{Simin Rashidi$^{a,b}$, and Bahram Golestani Eimani$^a$}
\address{$^a$Department of Biology, Faculty of Science, Islamic Azad University, Urmia, Iran.}
\address{$^b$(Current Affiliation) School of Medicine and Dentistry, Gold Coast Campus, Griffith University, Gold Coast, QLD 2222,
	Australia.}

\begin{abstract}
	
Due to the concerns of the society about the increase of antibiotic resistant infections, many studies and research have been done on nanoparticles and applications of nano-biotechnology. Zirconium Oxide ($\text{ZrO}_{2}$) in which called zirconia, is a white oxide of zirconium metal that its diameter is 20 nm. The colloidal size of these particles is often smaller than bacterial and eukaryotic cells. The main intention of this paper is to investigate the effect of different doses of $\text{ZrO}_{2}$ NPs on the sequence’s changes for the \textit{Escherichia coli} (\textit{E. coli}) genome. At the first step, \textit{E. coli} was cultured in eosin methylene blue agar and brain heart broth (BHB) mediums, respectively. Then, bacteria were treated with $\text{ZrO}_{2}$ NPs at concentrations of 100, 250, and 350 $\mu$g/ml. After treatment, the growth of bacteria was evaluated by utilizing spectrophotometry at 600 nm after incubation times including 2, 4, 6, 8, and 24 hours at 37 $^{\circ}$C. At the second step, the extraction of DNA was performed by using control and treated samples. Then, the changes in the sequence of bacterial genome were investigated using RAPD markers. Finally, NTSYS-PC platform was employed in order to analyze of the results extracted by electrophoresis of products on agarose gel. In this paper, it was observed that $\text{ZrO}_{2}$ NPs can inhibit the growth of bacteria at concentrations of 250 and 350 $\mu$g/ml after 8 hours of treatment. It was also found that the $\text{ZrO}_{2}$ NPs at different concentrations have not changed the genome sequence of \textit{E. coli}. Furthermore, it was concluded that the $\text{ZrO}_{2}$ NPs with the concentration of 350 $\mu$g/ml had the highest inhibitory properties without significant changing in the genomic sequence of \textit{E. coli}.
\end{abstract}


\end{frontmatter}


\section{Introduction}\label{sec1}

\textit{E. coli} is a gram-negative bacillus, oxidase and catalase-negative, facultative anaerobic, and non-spore forming. This bacteria decarboxylates the amino acid lysine and uses citrate as the only source of carbon. These bacteria are commonly found in the intestines of warm-blooded organisms and prevent the establishment of pathogenic bacteria in the intestine. \textit{E. coli} is an opportunistic organism that causes eye infections, urinary tract infection, pneumonia, meningitis, and septicemia for the human. Different strains of~\textit{E. coli} show diverse and complex pathogenic mechanisms in the occurrence of diarrhea. One of these pathogenic mechanisms is the production of various enterotoxins in which some of these enterotoxins cause pathogenesis in humans as well as some of them cause pathogenesis in animals~\cite{R1,R2}. It is difficult to control the spread of pathogenic bacteria because antibiotic resistance is a global problem between bacteria, leading to the spread of resistance among other bacteria and the emergence of more resistant strains. The use of NPs is a promising strategy against antibiotic resistance. In general, NPs are mainly categorized in two groups consisting of metallic and non-metallic from where metal NPs have significant antimicrobial properties on gram-positive and gram-negative bacteria~\cite{R3,R4}. NPs have some special properties that have been considered by scientists in recent years~\cite{R5}. NPs depict unique properties because of their size and surface to volume ratio in which the smaller size of the nanoparticle leads the greater the antibacterial activity as well as the high surface to volume ratio causes the NPs to interact with the microbial membrane~\cite{R6,R7,R8,R9}. NPs are widely used in different applications such as food processing and packaging, disinfection of textiles and equipment, and in the biosensors for rapid detecting of contaminants. In addition, NPs are utilized for therapeutic processes in biological and pharmaceutical research activities due to their higher potential as well as many studies have shown that NPs can have negative effects on the growth of microorganisms~\cite{R10,R11}.

There are many antibacterial mechanisms for metal NPs that are disruption of energy transduction, generating reactive oxygen species (ROS), ion release, biomolecular damage, and electrostatic reaction between NPs and bacterial membranes. The interaction of NPs with macromolecules are based on the difference between the negative charge of microorganisms and the positive charge of NPs and this difference leads to junction of NPs with cell surface and consequently cause cell death. Another antimicrobial mechanisms of metal NPs is that they perforate the bacterial cell wall and accordingly penetrate into the cell~\cite{R12,R13}. $\text{ZrO}_{2}$ NPs have been considered due to their better electrical and mechanical characteristics, high dielectric constant and wide band gap as well as they are extensively used in dental implants and building bone and hair tissue~\cite{R14}. Moreover, $\text{ZrO}_{2}$ is used to remove phosphate from contaminated water~\cite{R15}. Additionally, the powders of $\text{ZrO}_{2}$ are employed in chromatography because $\text{ZrO}_{2}$ is used as a support surface for the separation of proteins and enzymes~\cite{R16}. These NPs affect the bacterium \textit{Streptococcus mutans} by using their hydrophobic surface and attach to the bacterium instead of forming a biofilm. It is noticeable that increasing the hydrophobic surface of $\text{ZrO}_{2}$ NPs increases the adhesion forces of these NPs to attach bacteria~\cite{R17}. $\text{ZrO}_{2}$ NPs are also used in biological and medicinal applications and have high biomedical, antimicrobial, and anti-fungal activities as well as significantly inhibit the growth of \textit{Rhizoctonia solani}~\cite{R18}. Furthermore, $\text{ZrO}_{2}$ NPs are utilized in agriculture such as germination of plants~\cite{R19} as well as they have anti-fungal and antibacterial properties against \textit{Candida albicans} and \textit{E. coli}, respectively~\cite{R20}.

Metal NPs may cause mutations in bacteria and antimicrobial activity by altering the base pair~\cite{R21}. By considering the mentioned challenges, in order to use of $\text{ZrO}_{2}$ as an antibacterial agent on microorganisms, it is needed to be examined effective antimicrobial concentrations and their genotoxic effect as a suitable agent instead of antibiotics. The contributions of this paper are listed in the following:

$\bullet$ The effect of $\text{ZrO}_{2}$ NPs on the bacterial genome has not been investigated by now. This paper investigates the effect of different doses of $\text{ZrO}_{2}$ NPs on the genome of \textit{E. coli} bacteria. 

$\bullet$ The genotoxic changes of DNA sequences are also explored using RAPD-PCR method. It is important to note that the RAPD-PCR primers are randomly attached to different sequences on DNA and randomly amplify fragments. This method is used to determine the genome differences between two groups of organisms in terms of genomic DNA sequence.

The outlines of this paper are as follows: Section~\ref{sec2} explains the material and methods. The results of the experiments are provided in Section~\ref{sec3}. In Section~\ref{sec4}, appropriate comparisons between the presented work with other literature are explained through discussion. Finally, Section~\ref{sec5} remarks the conclusion including future directions.

\section{Material and Methods}\label{sec2}

This section presents the details of employed materials and methods. The Bacterial culture and the required conditions, treatment of bacteria with $\text{ZrO}_{2}$ NPs, and DNA extraction and RAPD-PCR along with proper evaluation of RAPD-PCR results as well as analyzes of electrophoresis data are investigated in this section.

\subsection{The Bacterial culture and the required conditions}\label{subsec2.1}

In this stage, \textit{E. coli} O157:H7 was characterized on eosin methylene blue agar. The bacteria were cultured in 7 ml of BHB medium and placed in a shaking incubator at 37 $^{\circ}$C with 200 rpm for 24 hours before treatment. The bacterial growth rate was measured by measuring the absorbance density (OD) of the culture medium at 600 nm wavelength~\cite{R15}.

\subsection{Treatment of bacteria with $\text{ZrO}_{2}$ NPs}\label{subsec2.2}

$\text{ZrO}_{2}$ NPs with 20 nm diameter were prepared and the characteristics of these NPs were analyzed by transmission electron microscopy (TEM). Then, to investigate the effect of $\text{ZrO}_{2}$ NPs on bacteria, phosphate-buffered saline with pH:7.4 were used as a solvent on the concentrations of 100, 250, and 350 $\mu$g/ml in order to make a solution. In the next step, after treatment with the mentioned three concentrations (i.e. 100, 250, and 350 $\mu$g/ml) and incubating at 37 $^{\circ}$C with 200 rpm, the OD of each test tubes were read at 600 nm wavelength.

\subsection{DNA extraction and RAPD-PCR}\label{subsec2.3}

DNA of the control and treated bacteria was extracted using DNA extraction kits (Genet Bio Cat: No.K-3000) according to the manufactures’ instructions. The quantity and quality of the extracted DNA were analyzed by spectrophotometry and electrophoresis on 1\% agarose gel, respectively. Furthermore, RAPD-PCR was used to investigate of $\text{ZrO}_{2}$ NPs’ effect on the bacterial genome. To this end, 10 random primers were used. It is worthy to note that the sequence and characteristics of these 10 primers are given in Table~\ref{Table1}. The 25 $\mu$l PCR reaction including 10.5 $\mu$l of deionized water, 1 $\mu$l of primer, 1 $\mu$l of extracted DNA, and 12.5 $\mu$l of the master mix. In addition, the temperature program of PCR is listed in Table~\ref{Table2}.

\begin{table}[t]
	\centering
	\caption{Nucleotide sequence of RAPD-PCR primers}
	\begin{tabular}{c c}
		\hline\hline
		\textbf{Primer name} & \textbf{Primer nucleotide sequence} \\[1pt]
		\hline
		OPR12 & ACAGGTGCGT \\ [1pt]
		OPS11 & AGTCGGTGG \\ [1pt]
		OPA11 & CAATCGCCGT \\[1pt]
		OPS13	& GTCGTTCCTG \\[1pt]
		OPQ14	& GGACGCTTCA \\[1pt]
		OPB08	& GGTGACGCAG \\[1pt]
		OPS03	& CAGAGGTCCC \\[1pt]
		OPT17	& TCTGGTGAGG \\[1pt]
		OPR11 &	GTAGCCGTCT \\ [1pt]
		OPS14	& AAAGGGGTCC \\ [1pt]
		\hline\hline
	\end{tabular}
	\label{Table1}
\end{table}

\begin{table}
	\centering
	\caption{The temperature program of PCR}
	\footnotesize
	\begin{tabular}{ l c c c c }
		\hline
		\multicolumn{5}{c}{Categories} \\
		\hline
		Cycle & Stage name & Temperature ($^{\circ}$C) & Time (sec) & Cycles’ number  \\
		\hline
		\multirow{1}{*}{First}
		& Initial denaturation & 95 & 300 & 1 \\
		\hline
		\multirow{3}{*}{Second}
		& Denaturation & 95 & 35 & 40  \\
		& Annealing & 30 &	45	& 40 \\
		& Extension & 72 & 45 & 40  \\
		\hline
		\multirow{1}{*}{Third}
		& Final extension & 72 & 420 & 1 \\
		\hline
	\end{tabular}
	\label{Table2}
\end{table}

\subsection{Evaluation of RAPD-PCR results}\label{subsec2.4}

After PCR reaction, for detecting the bands, 10 $\mu$l of PCR products were electrophoresed on 2\% agarose gel (26×14 cm) containing red safe in TBE buffer (1x) for 4 hours at 120 V. Furthermore, the DNA ladder marker was used at 100 to 1500 bp in order to determine the size of the product.

\begin{figure}[th]
	\centering
	\includegraphics[width=12cm]{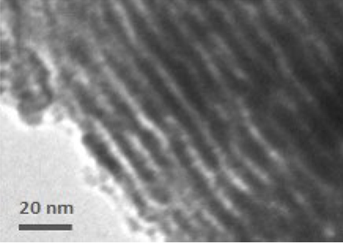}
	\caption{TEM image of $\text{ZrO}_{2}$ NPs} 
	\label{Figure1}
\end{figure}

\begin{figure}[th]
	\centering
	\includegraphics[width=12cm]{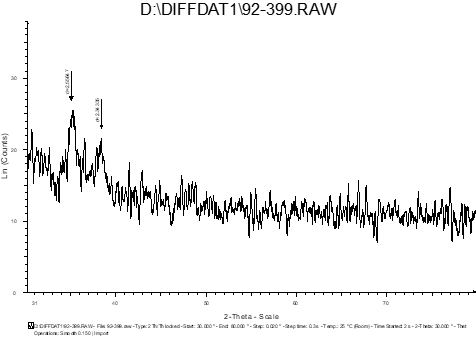}
	\caption{XRD analysis results for $\text{ZrO}_{2}$ NPs} 
	\label{Figure2}
\end{figure}

\subsection{Analyzes of electrophoresis data}\label{subsec2.5}

The bands on the gel were based on the existence and absence with 1 and 0, respectively. Then the data were molecularly analyzed in NTSYS-PC Dic platform and the dendrogram was plotted by (Un weighted Pair Group Mean Average) UPGMA method.

\section{Results}\label{sec3}

This section explains the results of transmission electron microscopy (TEM) and X-ray diffraction (XRD) of $\text{ZrO}_{2}$ NPs as well as the antimicrobial properties of these NPs are evaluated. Furthermore, RAPD-PCR product analyzes and the data analysis based on RAPD-PCR are investigated.

\subsection{Results for the outputs of TEM and XRD}\label{subsec3.1}

To verify the accuracy of $\text{ZrO}_{2}$ NPs, TEM is used. The extracted results by TEM from $\text{ZrO}_{2}$ NPs and XRD analysis are depicted in Figs. 1 and 2, respectively. In these figures, $\text{ZrO}_{2}$ NPs are seen in black spots and are stabilized in silica pores. In addition, zirconium NPs are less than 20 nm in size as shown in Fig.~\ref{Figure1}. In Fig.~\ref{Figure2}, XRD analysis confirms the presence of $\text{ZrO}_{2}$ NPs as well as 2-Theta consisting of 36 and 38 are related to $\text{ZrO}_{2}$ NPs.

\subsection{Evaluation of antimicrobial properties of $\text{ZrO}_{2}$ NPs}

\begin{figure}
	\centering
	\includegraphics[width=12cm]{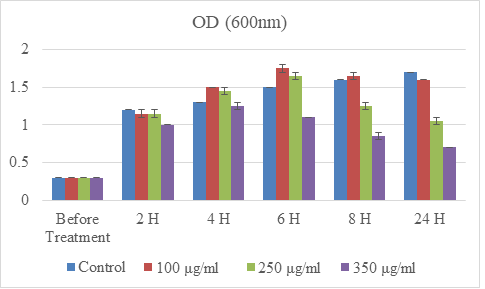}
	\caption{Evaluation of OD changes in proportion to changes in $\text{ZrO}_{2}$ nanoparticle concentrations at different times after treatment} 
	\label{Figure3}
\end{figure}

Table~\ref{Table3} summarizes the results of the antimicrobial activity of $\text{ZrO}_{2}$ NPs on \textit{E. coli} bacteria. As indicated in Table~\ref{Table3}, $\text{ZrO}_{2}$ NPs at the concentrations of 250 and 350 $\mu$g/ml inhibit the growth of \textit{E. coli} after 8 hours. Also, Fig.~\ref{Figure3} shows the rate of change of OD in proportion to the changes in the concentration of $\text{ZrO}_{2}$ NPs at different times after treatment. Fig.~\ref{Figure3} plots the changes in OD at different time intervals in proportion to changes in the concentrations of $\text{ZrO}_{2}$ NPs.

\begin{table}
	\centering
	\caption{Antimicrobial activity of $\text{ZrO}_{2}$ NPs}
	\footnotesize
	\begin{tabular}{ l c c c c c c c }
		\hline
		\multicolumn{8}{c}{Categories} \\
		\hline
		Sample & \makecell{Tube \\ number} & \makecell{Before \\ treatment} & \makecell{2 hours after \\ treatment} & \makecell{4 hours after \\ treatment} & \makecell{6 hours after \\ treatment} & \makecell{8 hours after \\ treatment} & \makecell{24 hours after \\ treatment}  \\
		\hline
		\multirow{1}{*}{Control}
		& 1 & 0.3 & 1.2 & 1.3 & 1.5 & 1.6 & 1.7 \\
		\hline
		\multirow{2}{*}{100 $\mu$g/ml}
		& 2 & 0.3 & 1.2	& 1.5 &	1.8  & 1.7 & 1.6 \\
		& 2 & 0.3 &	1.1	& 1.5 &	1.7	 & 1.6 & 1.6 \\
		\hline
		\multirow{2}{*}{250 $\mu$g/ml}
		& 3 & 0.3 &	1.2	& 1.5 &	1.6 & 1.2 &	1.1 \\
		& 3 & 0.3 &	1.1	& 1.4 & 1.7	& 1.3 &	1.0 \\
		\hline
		\multirow{2}{*}{350 $\mu$g/ml}
		& 4 & 0.3 &	1.0 &	1.3 & 1.1 &	0.8	& 0.7\\
		& 4 & 0.3 &	1.0 &	1.2 & 1.1 &	0.9 & 0.7 \\
		\hline
	\end{tabular}
	\label{Table3}
\end{table}

\subsection{RAPD-PCR product analyzes}\label{subsec3.3}

\begin{figure}[h]
	\centering
	\includegraphics[width=12cm]{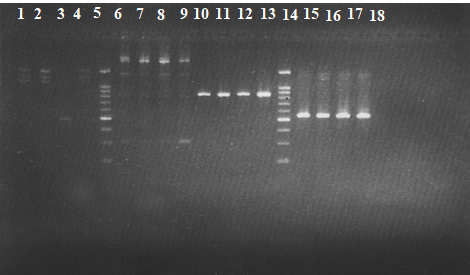}
	\caption{The electrophoresis results of RAPD-PCR product on agarose gel;~\newline Columns 1 to 4: Primer OPQ14, Column 5: Marker, Columns 6 to 9: Primer OPS13, Columns 10 to 13: Primer OPS14, Column 14: Marker, Columns 15 to 18: Primer OPS11} 
	\label{Figure4}
\end{figure}

\begin{figure}[h]
	\centering
	\includegraphics[width=12cm]{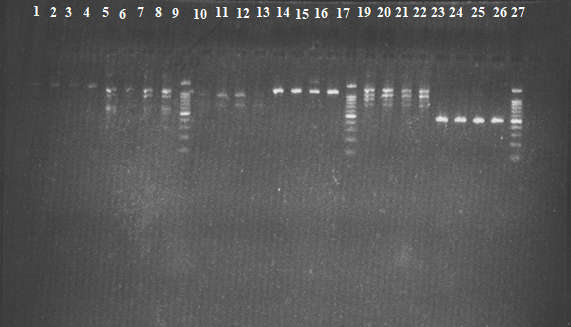}
	\caption{The electrophoresis results of RAPD-PCR product on agarose gel;~\newline Columns 1 to 4: Primer OPS03, Columns 5 to 8: Primer OPA11, Column 9: Marker, Columns 10 to 13: Primer OPB08, Columns 14 to 17: Primer OPA11, Column 18: Marker, Columns 19 to 22: Primer OPR12, Columns 23 to 26: Primer OPT17, Column 27: Marker} 
	\label{Figure5}
\end{figure}

The results of electrophoresis on RAPD-PCR products with 10 primers on 2\% (26×14 cm) agarose gel are illustrated in Figs. 4 and 5. The results for the bands were respectively scored based on their existence or absence in the samples with 1 or 0 in which are shown in Tables 4 and 5. Table~\ref{Table4} indicates the results for the bands extracted from Fig.~\ref{Figure4} as well as Table~\ref{Table5} shows the results for the bands extracted from Fig.~\ref{Figure5}. By considering Tables 4 and 5, the conclusion is based on the differences in the bands that each primer forms with the control and treated samples. In addition, the results for the primers are expressing in detail as the following terms:

$\bullet$ Primer OPQ14: There are 4 bands on the gel. It is worthy to note that all these 4 bands have difference between the control and treated groups. 

$\bullet$ Primer OPS13: There are 3 bands on the gel. For this primer, there is no difference between the control and treated groups. 

$\bullet$ Primers OPS11, OPS03, OPT17, and OPS14: For these primers, there is only 1 band on the gel. Furthermore, there are no difference between the control and treated groups. 

$\bullet$ Primer OPA11: There are 4 bands on the gel. It is important to note that 2 of these 4 bands have difference between the control and treated groups. 

$\bullet$ Primer OPB08: There are 2 bands on the gel. Also, all these bands have difference between the control and treated groups. 

$\bullet$ Primer OPR12: There are 5 bands on the gel. Moreover, there is no difference between the control and treated groups. 

$\bullet$ Primer OPS13: There are 3 bands on the gel. Again, there is no difference between the control and treated groups. 


\begin{table}[th]
	\centering
	\caption{Banding results related to the primers of Fig.~\ref{Figure4}}
	\footnotesize
	\begin{tabular}{ l c c c c c }
		\hline
		\multicolumn{6}{c}{Categories} \\
		\hline
		Primer & Band (bp) & C & $\text{T}_{1}$ & $\text{T}_{2}$ & $\text{T}_{3}$  \\
		\hline
		\multirow{4}{*}{OPQ14}
		& 1500 & 1 & 1 & 0 & 1 \\
		& 1400 & 1 & 1 & 0 & 1 \\
		& 1300 & 1 & 1 & 0 & 1 \\
		& 500 & 0 & 0 & 1 & 0 \\
		\hline
        \multirow{3}{*}{OPS13}
        & 1700 & 1 & 1 & 1 & 1 \\
        & 1400 & 1 & 1 & 1 & 1 \\
        & 300 & 1 & 1 & 0 & 1 \\
		\hline
		\multirow{1}{*}{OPS14}
		& 850 & 1 & 1 & 1 & 1 \\
		\hline
		\multirow{1}{*}{OPS11}
		& 550 & 1 & 1 & 1 & 1 \\
		\hline
	\end{tabular}
	\label{Table4}
\end{table}

\begin{table}[thb]
	\centering
	\caption{Banding results related to the primers of Fig.~\ref{Figure5}}
	\footnotesize
	\begin{tabular}{ l c c c c c }
		\hline
		\multicolumn{6}{c}{Categories} \\
		\hline
		Primer & Band (bp) & C & $\text{T}_{1}$ & $\text{T}_{2}$ & $\text{T}_{3}$  \\
		\hline
		\multirow{1}{*}{OPS03}
		& 1300 & 1 & 1 & 1 & 1 \\
		\hline
		\multirow{4}{*}{OPA11}
		& 1500 & 1 & 1 & 1 & 1 \\
		& 1300 & 1 & 1 & 1 & 1  \\
		& 950 & 1 & 0 & 1 & 1  \\
		& 600 & 1 & 0 & 1 & 1  \\
		\hline
		\multirow{2}{*}{OPB08}
		& 1000 & 1 & 1 & 1 & 0 \\
		& 700 & 1 & 0 & 1 & 0  \\
		\hline
		\multirow{2}{*}{OPR11}
		& 1600 & 0 & 0 & 1 & 0 \\
		& 1300 & 1 & 1 & 1 & 1 \\
		\hline
		\multirow{5}{*}{OPR12}
		& 1400 & 1 & 1 & 1 & 1 \\
		& 1100 & 1 & 1 & 1 & 1 \\
		& 1000 & 1 & 1 & 1 & 1 \\
		& 900 & 1 & 1 & 1 & 1 \\
		& 800 & 1 & 1 & 1 & 0 \\
		\hline
		\multirow{1}{*}{OPT17}
		& 500 & 1 & 1 & 1 & 1 \\
		\hline
	\end{tabular}
	\label{Table5}
\end{table}

\subsection{Analyzis based on RAPD data}\label{subsec3.4}

To analyzis the results based on RAPD data, the NTSYS-PC platform is employed. For this aim, the similarity matrix is calculated for the control and treated samples based on Dic method. The results of similiraty matrix are shown in Table~\ref{Table6} and indicate the genetic distance between the samples are vary from 0.82 to 1.000. It is noticale that the closer numbers to 1 means the greater genetic similarity among the samples. Reffering to the results from Table~\ref{Table6}, the genome of the treated bacteria is very similar to the control bacteria.

\begin{table}[b]
	\centering
	\caption{Similarity matrix for the control and treated samples}
	\footnotesize
	\begin{tabular}{ l c c c c c }
		\hline
		\multicolumn{6}{c}{Dice (Czekanowski or Sorenson) Measure} \\
		\hline
		 & & 1:ctr &	2:t100	& 3:t250 &	4:t350  \\
		\hline
		\multirow{4}{*}{Case}
	     & 1:ctr & 1.0000000  & & &  \\
		 & 2:t100	& 0.9900000	& 1.0000000  & & \\
		 & 3:t250 &	0.9268293 &	0.9368930 &	1.0000000 & \\
		& 4:t350  &	0.9090909 &	0.9091337 &	0.8292683 &	1.0000000 \\
		\hline
		\multirow{1}{*}{This is a similarity matrix}
		& &  &  &  &  \\
		\hline
	\end{tabular}
	\label{Table6}
\end{table}

\begin{figure}
	\centering
	\includegraphics[width=12cm]{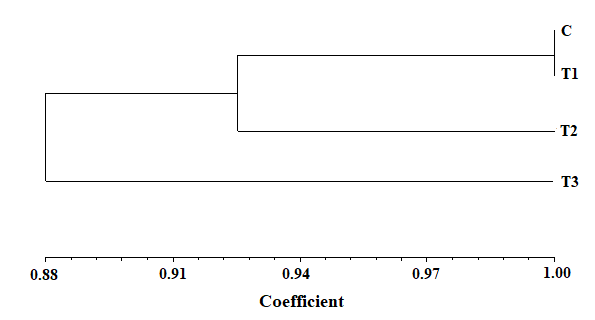}
	\caption{The extracted dendrogram from analyzes based on RAPD test using the UPGMA method} 
	\label{Figure6}
\end{figure}

Fig.~\ref{Figure6} indicates the extracted dendrogram from NTSYS-PC platform based on RAPD test analyzes using UPGMA method. The dendrogram shown in Fig.~\ref{Figure6} has drawn to evaluate the genetic diversity of treated samples with $\text{ZrO}_{2}$ NPs. According to the Fig.~\ref{Figure6}, it is observed that $\text{ZrO}_{2}$ NPs in treatment 3 with the concentration of 350 $\mu$g/ml has the highest inhibitory effect without significant changes in the bacterial genome.

\section{Discussion}\label{sec4}

Excessive use of antibiotics has made bacteria resistant to antibiotics. To resolve this problem, nanotechnology has been considered to combat microbial resistance. Most research has suggested that the use of NPs is a way to remove antibiotic-resistant bacteria~\cite{R22,R23}. The effect of NPs on the genomes of bacteria and other environmental organisms has great importance. In this study, specific concentrations of $\text{ZrO}_{2}$ NPs have been used to treat bacteria in order to evaluate their antimicrobial and genomic effect.

The results of the present study show that $\text{ZrO}_{2}$ NPs have good antimicrobial effect at doses of 250 and 350 $\mu$g/ml and can almost inhibit the growth of \textit{E. coli}. Jangra et al. Showed in their study that zirconia only inhibits \textit{E. coli}, while Zr(IV) crystals have inhibitory activity against gram-positive and gram-negative bacteria and fungal strains. These researchers also showed that the antimicrobial activity of zirconia and Zr(IV) crystals occurs through surface atoms~\cite{R24}. Gowri et al. found that $\text{ZrO}_{2}$ NPs had antimicrobial activity against \textit{E. coli} and \textit{Staphylococcus aureus}. In addition, these NPs have anti-fungal properties against \textit{Candida albicans} and \textit{Aspergillus niger}~\cite{R20}. However, the studies in~\cite{R20,R24} have not investigated the genotoxic effect of $\text{ZrO}_{2}$ NPs on the bacterial genome. Atalay et al. in~\cite{R25} examined the effect of $\text{ZrO}_{2}$ NPs on eukaryotic cells such as mouse fibroblasts and they have stated that $\text{ZrO}_{2}$ NPs cause damage and induce apoptosis in mouse fibroblast cells~\cite{R25}. In 2020, Khan et al. studied surface changes in $\text{ZrO}_{2}$ NPs and bonded glutamic acid to $\text{ZrO}_{2}$ NPs during their experiments~\cite{R26}. In~\cite{R26}, it is stated that when glutamic acid is bonded to $\text{ZrO}_{2}$ NPs, antimicrobial property is effectively enhanced compared to pristine $\text{ZrO}_{2}$ NPs. Aziz et al. in~\cite{R27} tested the antibacterial property of $\text{ZrO}_{2}$ NPs combined with ceftazidime on \textit{Klebsiella pneumoniae}. In~\cite{R27}, it is concluded that the antimicrobial property of ceftazidime combined with $\text{ZrO}_{2}$ NPs is higher than ceftazidime and $\text{ZrO}_{2}$ NPs. Furthermore, the combination of $\text{ZrO}_{2}$ NPs with ceftazidime is a useful method for the treatment of infections caused by \textit{Klebsiella pneumoniae}~\cite{R27}. Derbalah et al. examined and tested the role of $\text{ZrO}_{2}$ NPs on plants~\cite{R18}. In~\cite{R18}, it is concluded that $\text{ZrO}_{2}$ NPs have higher antifungal activity and significantly inhibit the growth of \textit{Rhizoctonia solani}. In addition,  these $\text{ZrO}_{2}$ NPs can control cucumber root rot disease, effectively~\cite{R18}. Jalil et al. have also investigated the antifungal property of $\text{ZrO}_{2}$ NPs in~\cite{R28} and stated that $\text{ZrO}_{2}$ NPs inhibit the growth of \textit{Fusarium graminearum} at a concentration of 200 mg/ml. In~\cite{R29}, Zhang et al. stated that the concentration of NPs plays a pivotal role compared to their size in view point of antimicrobial effect. It is concluded in~\cite{R29} that with increasing the concentration of NPs, the antimicrobial activity increases which is consistent with the results of the this paper. In 2015, Gawri et al. studied the antimicrobial properties of $\text{ZrO}_{2}$ NPs versus \textit{E. coli} and \textit{Staphylococcus aureus}~\cite{R30}. The results extracted from~\cite{R30} show that $\text{ZrO}_{2}$ NPs are effectively inhibit \textit{E. coli} than \textit{Staphylococcus aureus} with the same concentration of $\text{ZrO}_{2}$ NPs. It is also important to note that the antimicrobial activity of metal oxide NPs such as silver, copper and zinc has been proven in many studies~\cite{R31}. Regarding to the mentioned issues, this paper proves that $\text{ZrO}_{2}$ NPs have antimicrobial activity at higher concentrations and after about 8 hours. Furthermore, it is concluded that $\text{ZrO}_{2}$ NPs require more time and concentration for their operation. It is worthy to note that these NPs cannot be a good choice for the cases in which time and concentration are important.

Another important issue which has been considered in this paper is the effect of NPs on the bacterial genome. There are different methods for the interaction between NPs and bacteria~\cite{R32}. One of these methods is that the NPs attach to the bacterial membrane and penetrate into the bacterial cell. In this method, bacterial membranes consist of sulfur-containing proteins in which NPs react not only with these proteins, but also with phosphorus-containing compounds including DNA. Another method is that the NPs attack to the respiratory chain which lead to cell death~\cite{R33,R34}. As another mechanism for the effect of NPs on the bacteria is that the metal oxide NPs make disturbance in transcription and translation~\cite{R35}. Moreover, these metal oxide NPs cause fragmentation of single-stranded DNA and affect gene expression as well as can also affect the ability of RNA polymerase to open the helix and make transcription. In general, any factor that damages DNA can cause that cell to die~\cite{R36}. In this paper, in order to study the genotoxic effect of $\text{ZrO}_{2}$ NPs, \textit{E. coli} O157:H7 has been used as a model for gram-negative bacteria. By considering this issue, 10 primers have been used in the RAPD-PCR reaction. By this reaction, the existence or absence of the bands happens which means changing in the DNA sequence by $\text{ZrO}_{2}$ NPs. It is noticeable that some of these primers do not identify the target sequences and subsequently the relevant fragment is not amplified. The difference between the bands observed in the treated and control samples expresses that the primer target sequences have undergone changes in the treated bacteria that indicating a direct or indirect mutation by NPs~\cite{R37,R38}. Lee et al. have reported that by disrupting the mechanism of DNA polymerase, the replication will also be disrupted~\cite{R39}. Disrupting of replication causes changes in DNA sequences that lead to differences in target sequences by RAPD primers~\cite{R40}. It can be concluded from~\cite{R39,R40} that NPs can disrupt genes which control the mechanisms of transcription and replication and also can affect RNA polymerase. However, this paper demonstrates that $\text{ZrO}_{2}$ NPs do not cause significant mutations in the genome sequences of \textit{E. coli} bacteria.

\section{Conclusion}\label{sec5}

In this paper, the effect of $\text{ZrO}_{2}$ NPs studied on \textit{E. coli}. The results show that the growth of \textit{E. coli} bacteria is inhibited after treatment with $\text{ZrO}_{2}$ NPs at concentrations of 250 and 350 $\mu$g/ml after 8 hours. Therefore, $\text{ZrO}_{2}$ NPs are able to enter the bacterial cell and induce inhibiting effect. Furthermore, since $\text{ZrO}_{2}$ NPs have not altered the genome of \textit{E. coli} at different concentrations, it might be possible to utilize these NPs for therapeutic goals in order to deliver drugs to human cells because these NPs do not cause significant mutations in healthy cells. In conclusion, $\text{ZrO}_{2}$ NPs with a concentration of 350 $\mu$g/ml have the highest inhibitory property without significant altering in the genomic DNA sequence of \textit{E. coli} bacteria. The synergistic effect of $\text{ZrO}_{2}$ NPs and nisin on \textit{E. coli} genome will be focused and investigated as a future work.

\bibliography{E_COLI}

\end{document}